\documentstyle[11pt,epsf,amstex,righttag]{article}
\addtocounter{secnumdepth}{1}
\newcommand{\x}{\text{\bf x}}

\newcommand{\ba}{\text{\bf a}}

\newcommand{\be}{\text{\bf e}}

\newcommand{\r}{\text{\bf r}}

\numberwithin{equation}{section}
\begin{document}

\thispagestyle{empty}
\title{
Universal fluctuations\\ in the support of the random walk$^1$
} 
\author{F. van Wijland and H.J. Hilhorst\\
Laboratoire de Physique Th\'eorique et Hautes Energies$^2$\\
B\^atiment 211\\
Universit\'e de Paris-Sud\\
91405 Orsay cedex, France\\}

\maketitle
\vspace{-1cm}
\begin{small}
\begin{abstract}
\noindent A random walk starts from the origin of a $d$-dimensional
lattice.  The occupation number 
$n(\x,t)$ equals unity if after $t$ steps site $\x$ has 
been visited by the walk, and zero otherwise.
We study translationally invariant sums $M(t)$ of observables
defined locally on the field of occupation numbers.
Examples are the number $S(t)$
of visited sites; the area $E(t)$ of the (appropriately defined) surface
of the set of visited sites; and,
in dimension $d=3$, the  
Euler index of this surface.
In $d\geq 3$, the {\it
averages} $\overline{M}(t)$ all
increase linearly with $t$ as $t\rightarrow\infty$. 
We show that in $d=3$, to leading order in an asymptotic
expansion in $t$, the {\it deviations} from average $\Delta M(t)\equiv
M(t)-\overline{M}(t)$ are, up to a normalization, all
{\it identical to a
single "universal" random variable}. This result resembles an earlier one in
dimension $d=2$; we show that this universality breaks down for $d>3$.

\end{abstract}
\end{small}
\vspace{1.5cm}
\noindent L.P.T.H.E. - ORSAY 96/96\\
\\\\\\\\{\small$^1$Dedicated to Bernard Jancovici on his 65th birthday.}\\
{\small$^2$Laboratoire associ\'e au Centre National de la
Recherche Scientifique - URA D0063}
\newpage
\section{Introduction} 
\label{secintroduction}

We consider the {\it simple random walk} on a
$d$-dimensional hypercubic lattice: The
walk starts at time $t = 0$ at the origin {\bf x} $=$ {\bf 0} and steps at 
$t = 1, 2, 3,\ldots$ with equal probability to one of the 
$2d$ neighboring lattice sites. The set of the sites that have
been visited after $t$ steps is called the {\it support} at
time $t$.  
It is actually the support of the field of {\it occupation numbers}
$n(\x,t)$ defined by
\begin{equation}
n(\x,t)=\left\{ \begin{array}{cl}
0 & \text{if {\bf x} has not yet been visited at time } t\\ 
1 & \text{otherwise}
\end{array} \right.
\label{defnxt}
\end{equation}
The main variable that characterizes the support is its total number of
sites $S(t)$. The probability distribution of this 
random variable is known
explicitly only in dimension $d=1$, although it is well characterized
\cite{LeGall} also in higher dimensions. Our interest here will be in
the connection 
between the statistical properties of $S(t)$ and those of other
variables characterizing the support. 
An example of the variables of interest is 
the ($d-1$)-dimensional surface area
$E(t)$ of the support, defined as the total number of nearest-neighbor bonds
connecting a site of the support with
a site not in it.
We may imagine to obtain $E(t)$ 
by inspecting all pairs of neighboring sites on the lattice and
retaining a contribution unity whenever a pattern
is found that consists of a visited and an unvisited site.
Expressions for $S(t)$ and 
$E(t)$ in terms of the occupation numbers are easily constructed.
For later convenience we shall express them instead in terms of the
complementary occupation numbers
\begin{equation}
m(\x,t)\equiv 1-n(\x,t)
\label{defmxt}
\end{equation}
This leads to
\begin{eqnarray}
S(t) &=& \sum_{\x}\, [ 1 - m(\x,t) ]\nonumber\\
E(t) &=& \sum_{\x}\sum_{\be}\, [m(\x,t)+m(\x+\be,t)-2m(\x,t)m(\x+\be,t)] 
\label{exprSm}
\end{eqnarray}
where {\bf e} runs through the $d$ basis
vectors of the lattice.

Both $S(t)$ and $E(t)$ count the total number of occurrences in space of
specific local patterns of visited and unvisited sites. 
The quantities inside the summation signs in Eq.\,(\ref{exprSm}) are the
{\it indicator functions} of these patterns.
Clearly observables can be constructed that count
the total number of local patterns of a great many other types. 
A pattern may consist, for example, of a "dimer" of two visited
nearest-neighbor sites
oriented along a specified axis; or of an unvisited site
having $2d$ visited neighbors; {\it etc.~etc.} We shall generically denote such
observables and their linear combinations by the symbol $M(t)$. 
The class of all $M(t)$ is defined precisely in Sec.\,\ref{general}.

There certainly exist many nontrivial
correlations among the random variables $M(t)$. Nevertheless, it
should be possible, in principle, to express them in 
terms of independently
fluctuating degrees of freedom of the support. 
Examination \cite{VanWijlandetal} of dimension $d=2$ has yielded
a most surprising result.
It appears that to leading order as $t\rightarrow\infty$ the deviations
from average $\Delta M(t) \equiv M(t)-\overline{M(t)}$ satisfy
\begin{equation}
\Delta M(t) \simeq b_M\, t(\log t)^{-k_M-1}\,\eta(t) \qquad (d=2)
\label{DMD2}
\end{equation}
where $\eta(t)$ is a random variable of zero average and unit variance, 
$b_M$ a proportionality constant, and $k_M$ a nonnegative integer 
called the {\it order} of $M$
(one has for example $k_S=0$ and $k_E=1$).
Eq.\,(\ref{DMD2}) says that in the large $t$ limit 
the whole class of random variables $M(t)$
reduces to just {\it a single fluctuating
degree of freedom} $\eta(t)$.

In the present work we ask how the above conclusion is
affected by the spatial dimensionality $d$.
We are not aware of any simple heuristic argument that leads to 
(\ref{DMD2}) and that would provide an indication whether or not a
similar reduction of degrees of freedom still holds in higher dimension.
It should be expected that as $d$ goes up, independently fluctuating
degrees of freedom will appear.

We show by explicit calculation that in dimension $d=3$, 
to leading order as $t\rightarrow\infty$,
the fluctuating degrees of freedom still 
collapse to a single one in a manner very
similar to $d=2$. Explicitly,
\begin{equation}
\Delta M(t) \simeq a_M\,(t\log t)^{\frac{1}{2}}\,\xi(t)\qquad(d=3)
\label{DMD3}
\end{equation}
where $\xi(t)$ is a random variable of zero average and unit variance,
and $a_M$ a proportionality constant. 
Eq.\,(\ref{DMD3}) holds in particular for the choice $M=S$. The
fluctuations $\Delta S(t)$ have 
been thoroughly studied and we know \cite{LeGall} therefore that the 
probability distribution of $\xi(t)$ is Gaussian.
The variables $\xi(t)$ in $d=3$ and $\eta(t)$ in $d=2$ are {\it
universal} in the sense that they do not depend upon $M$.
Differences between $d=3$ and $d=2$ have to do,
among other things, with the appearance of an "order" $k_M$ in $d=2$ but
not in $d=3$. Also, $\eta(t)$ is known to be non-Gaussian.

The way to derive the three-dimensional result
is closely analogous to what was done in two dimensions. We calculate
the correlation of two
observables $M(t)$ and $M'(t)$ and find 
\begin{equation}
\overline{\Delta M(t)\Delta M'(t)} \simeq  a_M a_{M'}\, t\log t + \cal{O}(t)
\qquad (d=3)
\label{var3D}
\end{equation}
where $a_M$ and $a_{M'}$ are known proportionality constants.
We then exploit the fact that the prefactor in Eq.\,(\ref{var3D})
factorizes into an $M$ and an $M'$ dependent part.
It follows that the normalized random variables
\begin{equation}
\xi_M(t)\equiv a_M^{-1}(t\log t)^{-1/2}\Delta M(t)
\label{defxiM}
\end{equation}
have $\overline{\xi_M(t)}=0$ and $\overline{\xi_M(t)\xi_{M'}(t)}=1$ for
all $M$ and $M'$. Therefore the averages and variances of all 
differences $\xi_M(t)-\xi_{M'}(t)$ vanish and
the $\xi_M(t)$ must all be identical to one and
the same random variable that we shall call $\xi(t)$; whence (\ref{DMD3}).

We show, again by explicit calculation, that in all dimensions $d>3$ the
correlation between two arbitrary observables $M$ and $M'$ has the form
\begin{equation}
\overline{\Delta M(t)\Delta M'(t)} \simeq  c_{MM'}\, t \qquad (d>3)
\label{DMD4}
\end{equation}
with a prefactor $c_{MM'}$ that no longer factorizes.
Hence for $d>3$ the fluctuations of the $M(t)$ cannot be described any
more by a single random variable.

Our principal result, described above, is of a general nature. In
Sec.\,\ref{secexamples} we use our calculational framework to study
three examples of fluctuating observables $M(t)$.
One of these is 
the {\it Euler index} $\chi(t)$
of the surface of the support of the three-dimensional random walk.
We recall that the Euler index of a closed
two-dimensional surface embedded in three-dimensional space (the surface
of a "Swiss cheese") is
\begin{equation}
\chi=2(1-H+C)
\label{Eulerindex}
\end{equation}
where $H$ is the number of handles of the surface and $C$ the number of
cavities enclosed by it. 
In Sec.\,\ref{secexamples}, for appropriately defined "handles" and
"cavities", we construct 
an explicit expression for $\chi(t)$ and show that it is a member of the
class of $M(t)$. We then determine its asymptotic time dependence and
its correlation with other observables. 

\section{General observables $M(t)$}
\label{general}

The general property or "observable" $M(t)$ that we shall deal with can
now be defined as follows. Let $A$ be a finite (typically small) subset
of lattice vectors. Then we write
\begin{equation}
\label{Generalform}
M(t) = \sum_A \mu_A M_A(t)
\label{definM}
\end{equation}
where the $\mu_A$ are numerical coefficients, where the sum is through
all subsets $A$ not equivalent by a translation, 
and where 
\begin{equation}
M_A(t) = \sum_{\x} m_{\x+A}(t)
\label{definMA}
\end{equation}
with
\begin{equation}
m_{\x+A}(t) =  \prod_{\ba\in A} m(\x+\ba,t)
\label{definmxA}
\end{equation}
Here $\x+A$ is a convenient notation for the set of lattice sites
$\{\x+\ba\,|\,\ba\in A\}$.
When $A=\emptyset$
the RHS of Eq.\,(\ref{definmxA}) should be assigned the value unity
and Eq.\,(\ref{}) shows that $M_{\emptyset}(t)$ is the total number of
lattice sites.
As an example one may check that $S(t)$ has 
$\mu_A = \pm 1$ for $A = \emptyset$ and $A =\{\text{\bf 0}\}$, respectively,
and $\mu_A = 0$ otherwise; and that $E(t)$ has $\mu_A=2d,\,-2$ for
$A=\{\text{\bf 0}\},\,\{\text{\bf 0},\be\}$, respectively, 
and $\mu_A=0$ otherwise.
We note that the expansion coefficients $\mu_A$ should satisfy the relation
$\sum_A \mu_A = 0$
if the sum on $\x$ in Eq.\,(\ref{definMA}) 
is to have a finite
value in the infinite lattice limit.

\section{Averages of observables}
\label{averages}

\subsection{Averages $\overline{M}(t)$}
\label{secsimple}

The expression (\ref{definmxA}) averaged on all random walk trajectories
represents the probability that the set $\x+A$ has not yet been visited
by the walk. It is not surprising, therefore, that the average
$\overline{M}(t)$ of interest to us is connected to the distribution of 
first passage times for that set. 
More precisely, we can express it in terms of the probability
$f_{\x+A}(\tau)$ that the walker's first visit to the set $\x+A$
takes place at time $\tau$.
Elementary manipulations lead to
the explicit relation  
\begin{equation}
\hat{\overline{M}}(z)=-\frac{1}{1-z}\sum_{A\neq\emptyset}\mu_A\hat{F}_A(z) 
\label{exprMbarz}
\end{equation}
in which
\begin{equation}
\hat{F}_A(z) = \sum_{\x}\hat{f}_{\x+A}(z)
\label{definFA}
\end{equation}
and where, here and below, any time dependent quantity $X(t)$ that
occurs will be associated with a generating function
$\hat{X}(z)=\sum_{t=0}^\infty z^t X(t)$.
The first passage time probabilities can be expressed in the random walk
Green function by standard methods. After some algebra 
\cite{VanWijlandetal} one finds 
\begin{equation}
\hat{F}_A(z) = \frac{1}{(1-z)}G_A^{-1}(z)
\label{solnsumF1}
\end{equation}
in which
\begin{equation}
G_A^{-1}(z) = \sum_{\ba,\ba'\in A}(\text{\bf G}_A^{-1})_{\ba\ba'}
\label{defGA}
\end{equation}
where {\bf G}$_A$ is the matrix of elements $\hat{G}(\ba-\ba',z)$
with $\ba$ and $\ba'$ restricted to the set $A$, 
and $G(\x,t)$ is 
the probability that a walker starting at the
origin occupies site $\x$ after $t$ steps.
For the later purpose of
discussing the small-$(1-z)$ expansion it will be convenient
to split $\hat{G}(\x,z)$ up according to
\begin{equation}
\hat{G}(\x,z)=\hat{G}(\text{\bf 0},z)-g(\x,z)
\label{GG0g}
\end{equation}
With the aid of a little algebra we find the analogous splitting 
\begin{equation}
G_A(z) = \hat{G}(\text{\bf 0},z) - g_A(z)
\label{solnsumF2}
\end{equation}
in which
\begin{equation}
g_A^{-1}(z)=\sum_{\ba,\ba'\in A}(\text{\bf g}_A^{-1})_{\ba\ba'}
\label{defgA}
\end{equation}
where {\bf g}$_A$ is the matrix of elements $g(\ba-\ba',z)$
with $\ba$ and $\ba'$ restricted to the set $A$. 
Using Eqs.\,(\ref{solnsumF2}) and (\ref{solnsumF1}) in
Eq.\,(\ref{exprMbarz}) and 
inverting that equation we obtain the
final result for the average of a 
general observable $M(t)$,
\begin{equation}
\overline{M}(t) = -\frac{1}{2\pi \text{i}}\oint
\frac{\text{d}z}{z^{t+1}}\frac{1}{(1-z)^2}
 \sum_{A\neq\emptyset} \mu_A \frac{1}{\hat{G}(\text{\bf 0},z)-g_A(z)}
\label{resultMtaverage}
\end{equation}
where the integral runs counterclockwise around the origin.
The coefficient $G_A(1)$ appears in the
literature \cite{Lawler} as the electrostatic capacity of the set $A$. 

\subsection{Averages $\overline{M(t)M'(t)}$}
\label{secproducts}

The calculation of the correlation $\overline{\Delta M(t)\Delta M'(t)}$
requires that we now consider the average of the product $M(t)M'(t)$.
The latter can be rewritten as 
\begin{equation}
M(t)M'(t)=\sum_{\r}\sum_{A,B}\mu_A\mu'_BM_{A\cup(\r+B)}(t)
\label{MtMt'}
\end{equation}
The observable inside the sum on {\bf r} is of the same
type as $M(t)$ and $M'(t)$, even if it is parametrically dependent
on {\bf r}. The average of (\ref{MtMt'}) can therefore be evaluated with
the aid of Eq.\,(\ref{resultMtaverage}). After separating the terms with
$A=\emptyset$ or $B=\emptyset$ from the others we find 
\begin{equation}
\overline{M(t)M'(t)} = -\frac{1}{2\pi\text{i}}\oint
\frac{\text{d}z}{z^{t+1}} \frac{1}{(1-z)^2}
\sum_{A,B\neq\emptyset}\mu_A\mu'_B\sum_{\r} I_{AB}(\text{\bf r},z)
\label{solnMM't}
\end{equation} 
in which
\begin{equation}
I_{AB}(\r,z)=\frac{1}{G_{A\cup(\r+B)}(z)}-
\frac{1}{G_A(z)}-\frac{1}{G_B(z)}
\label{IABrz}
\end{equation}
This expression involves 
$G_{A\cup(\r+B)}(z)$, a quantity whose {\bf r} dependence needs to be
rendered more explicit if the sum on {\bf r} in Eq.\,(\ref{solnMM't}) 
is to be performed.  
This we are able to achieve only in the scaling limit
$r\rightarrow\infty$ and $z\rightarrow 1$ with $\xi^2\equiv 2dr^2(1-z)$
fixed. In that
limit a calculation analogous to that of Ref.\,\cite{VanWijlandetal}
for the two-dimensional case leads to
\begin{equation}
\frac{1}{G_{A\cup(\r+B)}(z)} \simeq
\frac{G_A(z)+G_B(z)-2\hat{G}(\r,z)}{G_A(z)G_B(z)-\hat{G}^2(\r,z)}
\label{solnFU}
\end{equation}
where
\begin{equation}
\hat{G}(\r,z) \simeq d^{d/2}\pi^{-d/2}(1-z)^{d/2-1}\xi^{1-d/2}
K_{d/2-1}(\xi)
\label{scalingG}
\end{equation}
The expressions of this section up through Eq.\,(\ref{IABrz}) 
are exact for all finite times $t$. 
Only in Eqs.\,(\ref{solnFU}) and (\ref{scalingG})
has the scaling limit been taken, is an infinite lattice
implied, and does the dimensionality of space
appear explicitly.

\section{Long time behavior}
\label{seclongtime}

\subsection{Averages $\overline{M}(t)$}
\label{secaverages2}
The long time behavior of
$\overline{M}(t)$ is determined by the behavior of
$\hat{\overline{M}}(z)$ 
in the vicinity of $z=1$. We need the small-$(1-z)$ expansion of 
$G_A(z)$, hence of 
$\hat{G}(\text{\bf 0},z)$ and $g_A(z)$. The expansion of
$\hat{G}(\text{\bf 0},z)$ is well-known 
\cite{MontrollWeiss,ZumofenBlumen} and reads
\begin{equation}
\hat{G}(\text{\bf 0},z)=\left\{\begin{array}{ll}
G_0-\frac{3^{3/2}}{2^{1/2}\pi}(1-z)^{1/2}+{\cal O}(1-z)&(d=3)\\
G_0-\frac{4}{\pi^2}(1-z)\ln\frac{1}{1-z}+c_4(1-z)+o\big((1-z)^{3/2}\big
)&(d=4)\\
G_0-c_d(1-z)+{\cal O}\big((1-z)^{3/2}\big)&(d\geq 5)
\end{array}\right.
\nonumber
\end{equation}
\vspace{-7mm}
\begin{equation}
\label{exprG0}
\end{equation}
where we have abbreviated $G_0\equiv\hat{G}(\text{\bf 0},1)$ and where
the $c_d$ with $d=4,5,\ldots$ are positive constants. Since $g(\x,z)$
is a solution of
\begin{equation}
\frac{1}{2d}\sum_{\mu=1}^d
\Big[g(\x+\be_\mu,z)-2g(\x,z)+g(\x-\be_\mu,z)\Big]=\frac{1}{
z}\delta_{\x,\text{\bf 0}}-\frac{1}{z}(1-z)\hat{G}(\text{\bf 0},z) 
\end{equation}
one may conclude that $g_A(z)$ defined in
Eq.\,(\ref{defgA}) has an expansion of the form  
\begin{equation}
g_A(z)=g_A(1)+(1-z)g_A^{(1)}+o(1-z)
\end{equation}as $z\rightarrow 1$, where $g_A^{(1)}$ is an $A$ dependent
numerical constant. The $(1-z)$-expansion of $\hat{\overline{M}}(z)$ 
therefore reads
\begin{equation}
\hat{\overline{M}}(z)=\left\{\begin{array}{l}
m_1(1-z)^{-2}+\frac{3^{3/2}}{2^{1/2}\pi}m_2(1-z)^{-3/2}
+{\cal O}\big((1-z)^{-1}\big)\\
m_1(1-z)^{-2}+\frac{4}{\pi^2}m_2(1-z)^{-1}\log\tfrac{1}{1-z}\\
\phantom{m_1(1-z)^{-2}}+(m_2-c_4m_2^{(1)})(1-z)^{-1}
+o\big((1-z)^{-1/2}\big)\\
m_1(1-z)^{-2}+(c_dm_2+m_2^{(1)})(1-z)^{-1}+{\cal
O}\big((1-z)^{-1/2}\big)\nonumber
\end{array}\right.
\end{equation}
\vspace{-7mm}
\begin{equation}
\end{equation}
in dimensions $d=3,\, d=4$, and $d\geq 5$, respectively; here the
coefficients $m_n$, $m_2^{(1)}$ are determined by the observable $M$
through the relations
\begin{equation}
m_n[M]=-\sum_{A\neq\emptyset}\frac{\mu_A}{G_A^n(1)},\qquad 
m_2^{(1)}[M]=-\sum_{A\neq\emptyset}\frac{\mu_A g_A^{(1)}}{G_A^2(1)}
\label{defmn}
\end{equation}
When no confusion can arise we shall not indicate their $M$ dependence
explicitly. For transforming back to the time domain we need the basic
expansions, for $s>0$, 
\begin{equation}\label{defIs}
\frac{1}{2\pi{\text i}}\oint \frac{{\text d}z}{z^{t+1}}\frac{1}{(1-z)^s}=
\frac{t^{s-1}}{\Gamma(s)}\Big[1+\frac{s^2-s}{2}\frac{1}{t}+{\cal
 O}\Big(\frac{1}{t^2}\Big)\Big]
\label{int1}
\end{equation}
\begin{equation}\label{defJ}
\frac{1}{2\pi{\text i}}\oint
\frac{{\text d}z}{z^{t+1}}\frac{1}{(1-z)^s}\ln\frac{1}{1-z}=
\frac{t^{s-1}}{\Gamma(s)}\Big[\log t 
-\psi(s)+{\cal O}\Big(\frac{\log t}{t}\Big)\Big]
\label{int2}
\end{equation}
where $\psi$ is the psi function. 
The first integral is easily found by writing the
integrand as a power series in $z$ and the second one follows from
it by differentiation with respect to $s$. The result is
\begin{equation}
{\overline{M}}(t)=\left\{\begin{array}{ll}
m_1 t+\frac{2^{1/2}3^{3/2}}{\pi^{3/2}}m_2 t^{1/2}+{\cal O}(1)&(d=3)\\
m_1 t+\frac{4}{\pi^2} m_2\ln
t+m_1+\frac{4\gamma}{\pi^2}m_2-c_4m_2+m_2^{(1)}+o(t^{-1/2})&(d=4)\\
m_1 t+m_1+c_d m_2+m_2^{(1)}+{\cal O}(t^{-1/2})&(d\geq 5)
\end{array}\right.\nonumber
\end{equation} 
\vspace{-7mm}
\begin{equation}
\label{resMt}
\end{equation}
where $\gamma$ denotes Euler's constant.
In Sec.\,\ref{secexamples} some of the coefficients appearing in
Eq.\,(\ref{resMt}) will be calculated for specific
examples of observables $M(t)$. 

\subsection{Fluctuations and correlations}
\label{secfluctuations}

In order to study the fluctuations and correlations of two observables
$M(t)$ and $M'(t)$ in the large time limit, we now have to 
evaluate the average (\ref{solnMM't}) in the limit
$t\rightarrow\infty$. 
The principal problem is to find the expansion of
$\sum_{\r}I_{AB}(\r,z)$ as $z\rightarrow 1$.
Whereas $G_A(z)$ and $G_B(z)$ remain finite
in that limit, Eq.\,(\ref{scalingG}) shows that in all 
$d>2$ the quantity $\hat{G}(\r,z)$ vanishes when $z\rightarrow
1$ (with $\xi$ fixed). This suggests that different leading powers of
$1-z$ are associated with different powers of $\hat{G}(\r,z)$. We
therefore express $I_{AB}$ in the form   
\begin{equation}
I_{AB}(\r,z)= I_1(z)\hat{G}(\r,z)+\cdots+
I_{n-1}(z)\hat{G}^{n-1}(\r,z)+f_{AB}^{(n)}(\r,z)
\label{splitIAB}
\end{equation}
in which we take for $I_k(z)$ the coefficient of $\hat{G}^k(\r,z)$ in a power
series expansion of the RHS of 
Eq.\,(\ref{IABrz}) in which for $1/G_{A\cup(\r+B)}(z)$ the RHS of 
Eq.\,(\ref{solnFU}) has been
substituted. Explicitly, 
\begin{eqnarray}
I_1(z) &=& -2/G_A(z)G_B(z)\nonumber\\
I_2(z) &=& (G_A(z)+G_B(z))/G_A^2(z)G_B^2(z)\nonumber\\
I_3(z) &=& -2/G_A^2(z)G_B^2(z)
\label{defIn}
\end{eqnarray}
Even though Eq.\,(\ref{solnFU}) holds only in the scaling limit,
Eq.\,(\ref{splitIAB})
is an exact identity if it is taken as the definition of the remainder
$f_{AB}^{(n)}$.
It is now possible to find the small $1-z$ expansion of
$\sum_{\r}I_{AB}(\r,z)$ in the following way. 
After substituting Eq.\,(\ref{splitIAB}) in Eq.\,(\ref{MtMt'})
we evaluate the sum on all space of the powers 
$\hat{G}^k(\r,z)$ for $k=1,2,\ldots$. For $k=1$ conservation of
probability trivially leads to $\sum_{\r}\hat{G}(\r,z)=(1-z)^{-1}$.
For $k=2$ one easily derives the general relation
\begin{equation}
\sum_{\r}\hat{G}^2(\r,z)=(1+z\frac{\text d}{{\text d}
z})\hat{G}(\text{\bf 0},z)
\label{}
\end{equation}
after which use can be made of Eq.\,(\ref{exprG0}). It follows that 
\begin{equation}
\sum_{\r}\hat{G}^2(\r,z)=
\left\{
\begin{array}{l}
\frac{3^{3/2}}{2^{3/2}\pi}(1-z)^{-1/2}+{\cal O}(1)\\
\frac{4}{\pi^2}\ln\frac{1}{1-z}+G_0-c_4-\frac{
4}{\pi^2}+{\cal O}\big((1-z)^{1/2}\big)
\end{array}
\right.
\label{sumG2}
\end{equation}
in dimensions $d=3$ and $d=4$, respectively. In dimensions $d>4$
this quantity does not diverge as $z\rightarrow 1$. Finally, for
$k=3$, we employ the scaling expression (\ref{scalingG}) for
$\hat{G}(\r,z)$. In dimension $d=3$ the resulting integral on $\xi$ diverges
logarithmically at the origin, and using the lower end cutoff $\xi\sim
r\sqrt{1-z}$ we find
\begin{eqnarray}
\sum_{\r}\hat{G}^3(\r,z)&=&\frac{3^3}{2\pi^2}\ln\frac{1}{1-z}+{\cal
O}(1) \qquad (d=3)
\label{sumG1}
\end{eqnarray}
In dimensions $d>3$ the sum with $k=3$ remains finite as
$z\rightarrow 1$.
The remainder of the work will be carried out separately for the 
three qualitatively different  cases $d=3$,\, $d=4$, and $d\geq 5$.

\subsubsection{Dimension $d=3$}

We employ Eq.\,(\ref{splitIAB}) with $n=4$. 
In the limit $z\rightarrow 1$ the sum $\sum_{\r}f_{AB}^{(4)}(\r,1)$ remains
finite.  Using the above expansions and also expanding $G_A(z)$ and
$G_B(z)$ for small $1-z$ we find for the average of
Eq.\,(\ref{solnMM't}) the expression
\begin{eqnarray}
\overline{M(t)M'(t)}&=& \frac{1}{2\pi{\text i}}\oint
\frac{{\text d}z}{z^{t+1}}\Big[
2m_1 m_1^\prime \frac{1}{(1-z)^3}\nonumber\\
& &+\frac{3^{5/2}}{2^{3/2}\pi} (m_1m'_2+m'_1m_2) \frac{1}{(1-z)^{5/2}}
\nonumber\\ 
& &+\frac{3^3}{2\pi^2} m_2m'_2 \frac{1}{(1-z)^2}\log\frac{1}{1-z}
+\ldots\Big]
\label{expanMM't}
\end{eqnarray}
in which the $m'_n$ are determined by $M'$ via Eq.\,(\ref{defmn}).
After carrying out the $z$ integral with the aid of the basic relations
of Eqs.\,(\ref{int1}) and (\ref{int2}) and subtracting the product
$\overline{M}(t)\overline{M'}(t)$ obtainable from
Eq.\,(\ref{resMt}) we are led to the final result
\begin{equation}
\overline{\Delta M(t)\Delta M'(t)}\simeq\frac{3^3}{2\pi^2}m_2m_2^\prime\; t\ln
t+ c_{MM'}\,t
\label{d3}
\end{equation}
where the constant $c_{MM'}$ is due to contributions from various
subleading terms that have not been displayed explicitly above.
Eq.\,(\ref{d3}) leads to the result announced in Eq.\,(\ref{var3D}) of the
Introduction, whose consequence is the {\it universality} embodied by 
Eq.\,(\ref{DMD3}).

\subsubsection{Dimensions $d=4$ and $d\geq 5$}

We employ Eq.\,(\ref{splitIAB}) with $n=3$. 
In the limit $z\rightarrow 1$ the sum $\sum_{\r}f_{AB}^{(3)}(\r,1)$ remains
finite. Proceeding as in the case of $d=3$ we obtain in
dimension $d=4$ the final
result
\begin{equation}
\begin{array}{l}
\overline{\Delta M(t)\Delta M'(t)}=\\
\Big[m_1m_1^\prime-G_0(m_1m_2^\prime+m_2m_1^\prime)+m_1
m_2^{(1)\prime}+m_2^{(1)}m_1^\prime-{\cal F}_{MM'}\Big]t+\ldots
\end{array}
\label{d4}
\end{equation}
where 
\begin{equation}
{\cal F}_{MM'}=
\sum_{A,B\neq\emptyset}\mu_A\mu_B^\prime\sum_{\r}f_{AB}^{(3)}(\r,1)
\end{equation}
The only difference that arises when $d\geq 5$\, is that not only
$\sum_{\r}f_{AB}^{(3)}(\r,1)$ but also
$\sum_{\r}\hat{G}^2(\r,1)$ remains finite as $z\rightarrow 1$. 
Straightforward calculation leads in a slightly different way to exactly
the same result (\ref{d4}).  
If we denote the coefficient of the term linear in
$t$ again by $c_{MM'}$, then
Eq.\,(\ref{d4}) constitutes the result announced in
Eq.\,(\ref{DMD4}).

\section{Examples}
\label{secexamples}

We apply the preceding theory to three examples. As a
preliminary we render more precise the concept of "surface of the
support". We imagine $\Bbb{R}^d$ partitioned into unit cubes
that surround the lattice sites. 
Visited (unvisited) unit cubes correspond to visited (unvisited) lattice
sites. The support of the walk then defines a  
visited volume whose $(d-1)$-dimensional 
surface will be called the {\it surface of the support}. 
In the second example we shall study the Euler index of the
surface of the support of the three-dimensional walk. 
In order for this topological invariant -- which involves the handles 
and the cavities of the support --
to be well defined, we adopt the convention 
that two visited unit cubes are connected as soon as they share 
a vertex (and hence {\it a fortiori} when they share an edge or a face).
Fig.\,1 shows an example of two connected unit cubes; their
connectedness may be made visible by an infinitesimal deformation of the
surface near the common vertex.

\subsection{Surface area $E(t)$ of the support}
From Eq.\,(\ref{resMt}) we deduce that the average surface area of
the support in dimension $d\geq 3$ behaves as 
\begin{equation}
\overline{E}(t)=m_1[E] t +o(t)
\end{equation}
In the particular case $d=3$ we have, more precisely,
\begin{equation}
\overline{E}(t)=m_1[E] t+
\frac{2^{1/2}3^{3/2}}{\pi^{3/2}}m_2[E]\sqrt{t}+{\cal O}(1) \qquad (d=3)
\end{equation}
From Eq.\,(\ref{defmn}) and the remarks below Eq.\,(\ref{definmxA}) it
follows that for all $d$
\begin{equation}
m_n[E]=2d[(G_0-\frac 12)^{-n}-G_0^{-n}]
\end{equation}
where we used that for $A=\{\text{\bf 0}\},\,\{\text{\bf 0},\be\}$ 
one has $G_A=G_0,\,G_0-\frac{1}{2}$, respectively. In $d=3$
we have $G_0=1.5164...$ (\cite{MontrollWeiss}) so that
\begin{equation}
m_1[E]=1.95,\qquad m_2[E]=3.20 \qquad (d=3)
\end{equation}
In dimension $d\gg 1$, using the expansion $G_0=1+(2d)^{-1}+{\cal
O}(d^{-2})$ one sees that $\overline{E}(t)\simeq [2d-3+{\cal
O}(d^{-1})]t$, which may also be found by more heuristic
arguments. 

\subsection{Euler index $\chi(t)$}
The Euler index of an arbitrary, not necessarily closed,
two-dimensional surface embedded in $\Bbb{R}^d$ (with $d\geq 2$) is 
\begin{equation}\label{Eulerindexexact}
\chi=2(1-H+C)-B
\end{equation}
where $H, C$, and $B$ are the number of handles, cavities (connected
components of the complement of the surface), and boundaries, respectively.

The support of the two-dimensional walk is a flat two-dimensional
surface with $H=C=0$ and therefore has Euler index $\chi=2-B$; the
number of boundaries $B$ is equal to one plus the
number of unvisited islands enclosed by the support. This quantity was
studied in Ref. \cite{VanWijlandetal}. 

Here we wish to study the Euler index of the surface of 
the support of the {\it three}-dimensional walk. 
For this closed ($B=0$) two-dimensional surface the Euler index
reduces to 
$\chi=2(1-H-C)$. The handles and cavities are nonlocal objects of varying 
sizes and shapes, and it is interesting that one should be able
to calculate the Euler index.
The reason that this is possible is that the Euler index can be
expressed, by means of the Gauss-Bonnet theorem, 
as an integral over the surface of a function of the local curvature.

We first need a discrete analog of the Gauss-Bonnet theorem, that is, an
expression for $\chi$ as the sum over space of a function defined
locally on a two-dimensional surface built out of
plaquettes. In order to find the contribution from the surface around a
specific vertex, we examine all $2^8$ patterns of occupation
numbers in the surrounding cubic volume of $2\times 2\times 2$ unit
cubes. 
The local contribution to the Euler index will be a weighted sum
of the indicator functions of these patterns.  
There are only 22 patterns not equivalent under symmetry operations, and
therefore 22 weights to be determined.
One can set up equations for these weights by requiring that the final
expression correctly yield the known result
for various surfaces having the topology of a
sphere ($\chi=2$) or of a torus ($\chi=0$). 
One finds, for example, that the pattern shown in Fig.\,1 has weight
$-3$. 

It is then a tedious but straightforward 
matter to cast the expression for $\chi$ in the
form of Eq.\,(\ref{Generalform}) by working out the products with factors
$1-m$, summing on $\x$, 
and rearranging terms. 
Since the resulting expression for $\chi$ may also be of use in other
contexts, we have listed the coefficients in Fig.\,2.  
The nonzero coefficients $\mu_A$ are the 18 entries of that table, {\it and}
those obtainable from them by rotations and reflections.
Also listed are
the values of the capacities $G_A(1)$, found numerically by determining the
required values of $\hat{G}(\x,1)$ and inverting the 18
matrices {\bf G}$_A$. These data then allow the large time behavior 
of the average $\overline{\chi}(t)$ to be calculated.
It has the general form of Eq.\,(\ref{resMt}),
\begin{equation}
\overline{\chi}(t)=m_1[\chi] t+
\frac{2^{1/2}3^{3/2}}{\pi^{3/2}}m_2[\chi]\sqrt{t}+{\cal O}(1) \qquad (d=3)
\label{exprchi}
\end{equation}
 with the numerical values
\begin{equation}
m_1[\chi]=-3.44\times 10^{-2}, \qquad m_2[\chi]=2.12\times 10^{-1}
\qquad (d=3)
\end{equation}
The negativity of $m_1[\chi]$ implies that the Euler index is dominated
by the number of handles. 

The Euler index is defined also for 
$(d-1)$-dimensional surfaces and can always be expressed as the integral
of a local function on this surface. We have found no general
way to obtain discrete analogs of these expressions when the surface
dimension is $d-1$, with $d\geq 4$, even though for a given dimension this is
in principle possible.

\subsection{Cavities}
A quantity of potential interest is the number of cavities of a given 
size, shape and
orientation. The indicator function of such a cavity is 
a single product of $m$'s and $(1-m)$'s and their total number is
obtained by summing that product on space. This number therefore belongs to
the class of obervables $M(t)$ 
and behaves for large times as shown in Eq.\,(\ref{resMt}).

Let $C_1(t)$ be the number of cavities consisting of a single site. We
determined the numerical values 
\begin{equation}\label{cavit1}
m_1[C_1]=7.92\times 10^{-4}, \qquad m_2[C_1]=-7.10\times 10^{-3} \qquad (d=3)
\end{equation}
These show the rarity of such
cavities: Their number increases at the approximate rate of one
every one thousand steps. Cavities larger than a single site will
certainly be still rarer, so that $C_1(t)\approx C(t)$. It follows that
$\chi(t)$ is approximately equal to minus twice the number $H(t)$ of
handles of the surface of the support.

\subsection{Correlations}
\label{seccorrelations}
We have not included in the above examples the number of sites $S(t)$ of
the support, as this quantity has been thoroughly studied in the
literature \cite{Weiss,Hughes}. It is now useful to observe 
that the average $\overline{S}(t)$ behaves, in $d=3$, according
to the general law (\ref{resMt}) with coefficients $m_1[S]$ and $m_2[S]$
that are easily determined via our general rules. One finds 
\begin{equation}
m_1[S]=0.659, \qquad m_2[S]=0.434 \qquad (d=3)
\label{mS}
\end{equation}
in agreement with the literature \cite{Hughes}.
Upon inserting the various coefficients $m_2$
determined above in
Eq.\,(\ref{d3}) we directly obtain the leading order behavior
of all cross-correlations between the quantities studied. 
It appears that the total surface area $E(t)$ of the
support is positively correlated with $S(t)$, and that (since
$C_1(t)\approx C(t)$) the numbers of
handles and of cavities, $H(t)$ and $C(t)$, are negatively
correlated with $S(t)$.


\section{Discussion}
\label{secdiscussion}

The asymptotic behavior as $t\rightarrow\infty$ of the average and
the variance of the total number $S$ of sites in the support is well
known in all dimensions. These known results are reproduced by the relations
of this paper if one sets $M=M'=S$.
In the dimensions $d=2$ and $d=3$ the variable $S$ is known to have anomalous 
fluctuations in the sense
that the rms deviation $\overline{\Delta
S^2}^{1/2}$ increases faster with $t$ 
than proportional to $\overline{S}^{1/2}(t)$.
The fluctuations in $S$ do not arise, therefore, as the sum of effectively
independent contributions along the random walk trajectory, but must be 
due to
a collective effect. It is easy to accept that similar 
anomalous fluctuations appear in an entire class of observables $M(t)$
related to the support, and in their correlations. What is surprising,
however, is that the $M(t)$ all fluctuate in strict proportionality to
one another. This proportionality holds 
both in dimension $d=3$ (a result of this work) and in $d=2$
(derived in Ref.\,\cite{VanWijlandetal}). In $d\geq 4$ 
the anomalous fluctuations and the rigorous proportionality of the
fluctuating observables disappear.

The number of sites $S(t)$ visited by a $d$-dimensional lattice random
walk has its analog, in continuum space, in the $d$-dimensional volume
("{\it Wiener sausage}") swept out during a time interval $t$ by a 
hypersphere of radius $b$ that performs Brownian motion.
The volume of the Wiener sausage has been a recurrent object of 
investigation (see {\it e.g.} Ref.\,\cite{Russes}), 
and the islands enclosed by it in $d=2$ have been considered by
mathematicians ({\it e.g.} recently by Werner \cite{Math1}). 
However, we are not aware 
of any results, in dimensions $d\geq 3$, on the total surface area of
the Wiener sausage, let alone
on such properties as the number of handles or cavities. We see here problems for future research.


\newpage
\noindent FIGURE CAPTIONS\\

\noindent {\bf Figure 1.} Example of a local configuration of $2\times 2\times 2$ sites that may occur in the support of the three-dimensional walk. Shaded (unshaded) unit cubes are centered on visited (unvisited) lattice sites. The volumes of shaded cubes are considered to be connected via the common vertex.\\\\\\

\noindent {\bf Figure 2.} The Euler index can be written as $\chi=\sum_A \mu_A M_A$, with the $M_A$ defined in the text. The first and fifth column list the subsets $A$ of lattice sites (heavy dots) that have nonzero $\mu_A$. The other nonzero $\mu_A$ correspond to the sets $A$ obtainable from those shown by rotations and/or reflections. The total number of distinct sets related to $A$ by such symmetry operations is $\sigma_A$. The coefficients $G_A(1)$ are the capacities needed in the calculation.
\newpage
\begin{figure}
\epsfxsize=\hsize
\epsfbox{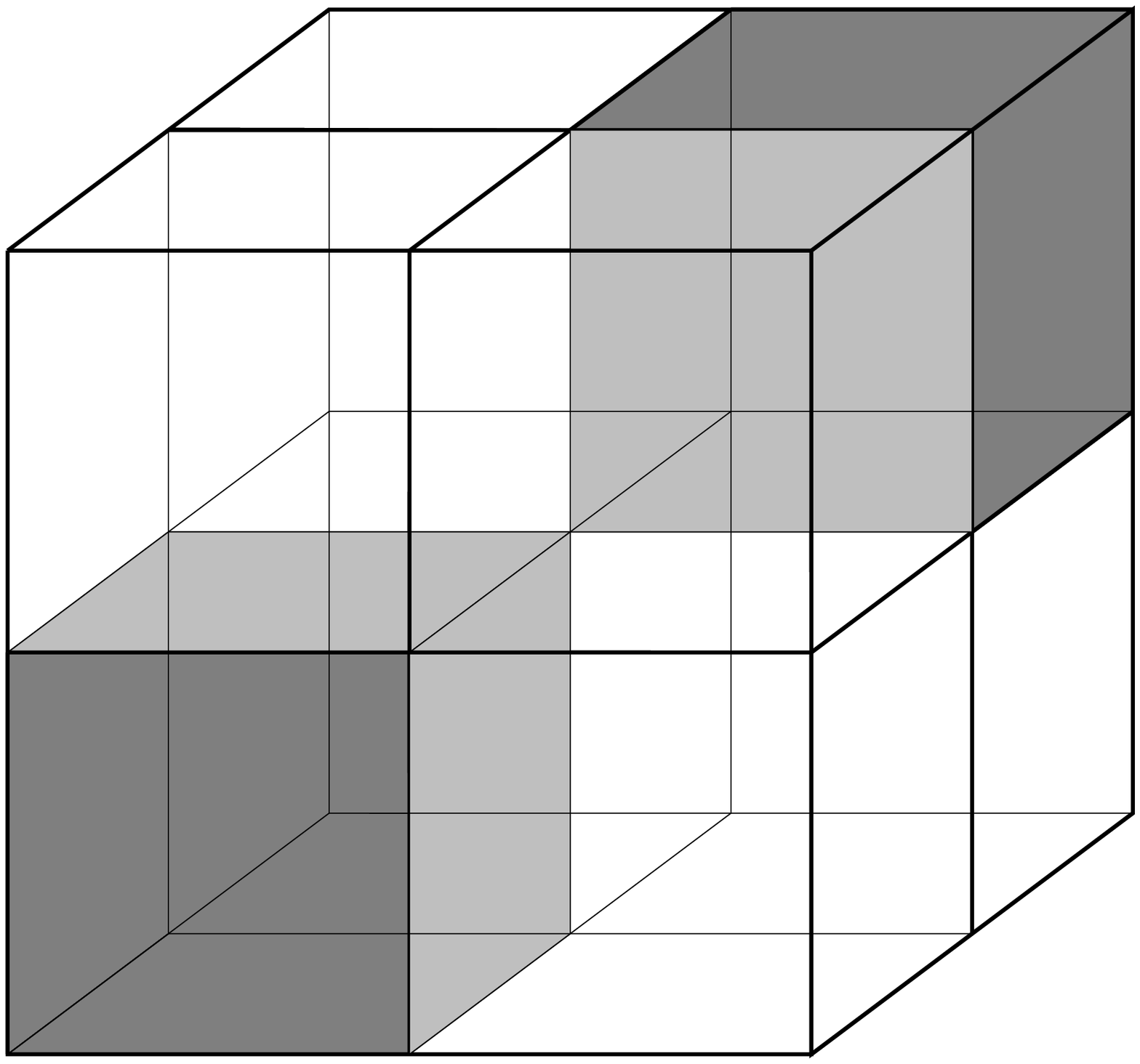}
\vskip 0.2in
\center{Figure 1}
\end{figure}
\newpage
\begin{figure}
\epsfxsize=\hsize
\epsfbox{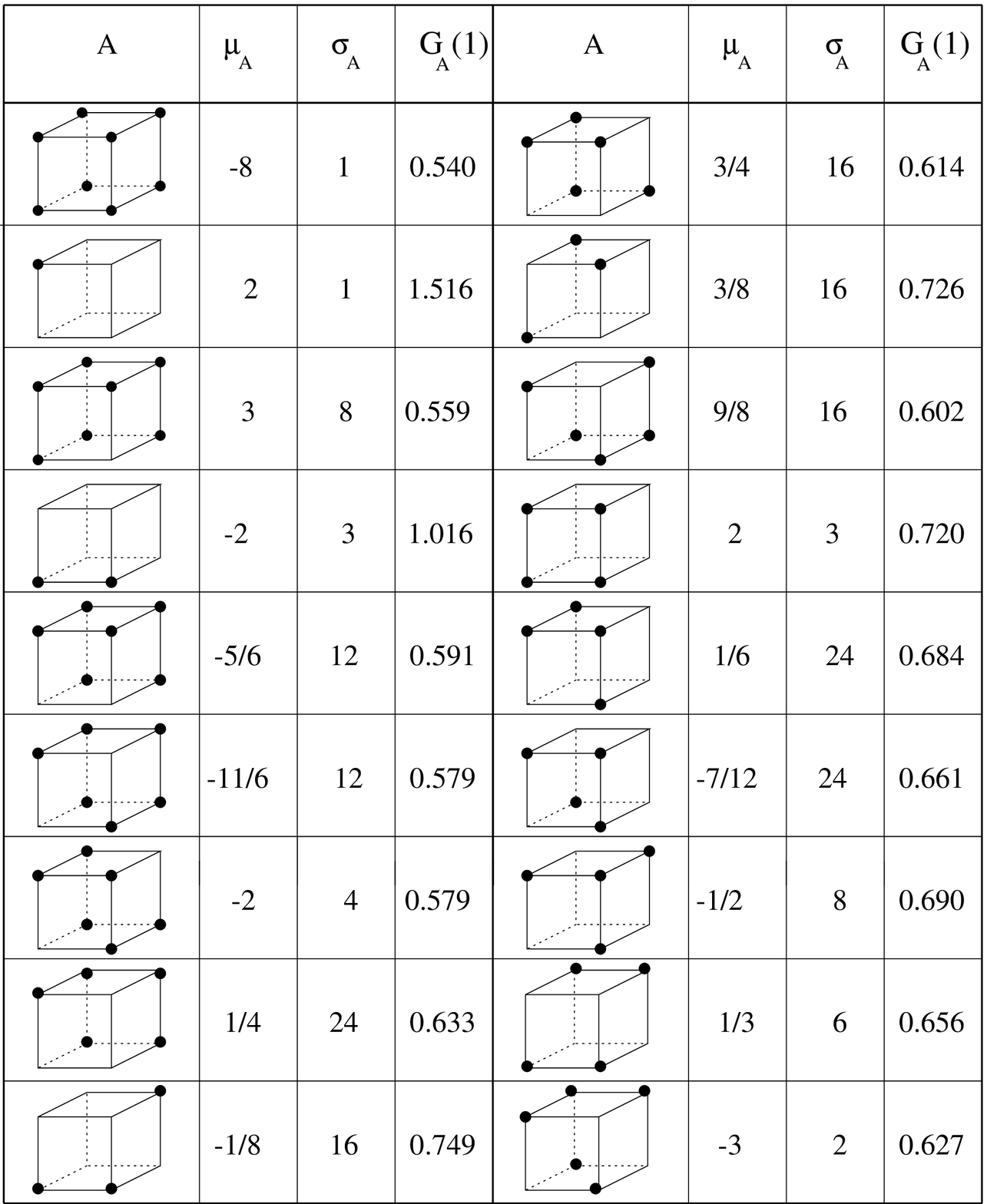}
\vskip 0.2in
\center{Figure 2}
\end{figure}

\end{document}